\documentclass{article}
\usepackage{ifpdf}
\ifpdf%
\usepackage{ae} 
\usepackage{aeguill}
\fi
\usepackage[backref=section]{hyperref}
\hypersetup{
    colorlinks,
     citecolor=blue,
     filecolor=black,
     linkcolor=cyan,
     urlcolor=red
}
\renewcommand*{\backref}[1]{}
\renewcommand*{\backrefalt}[4]{%
    \ifcase#1%
          \or\mbox{[Cited \S~#2.]}%
          \else\mbox{[Cited \S~#2.]}%
    \fi%
    }
\usepackage{amsmath}
\allowdisplaybreaks[4]
\def\matrixsize#1#2{{{#1}\times{#2}}}
\def\FMMA#1#2#3#4{{(#1\times{#2}\times{#3}\,;#4)}}
\newtheorem{mcounter}{mcounter}
\newtheorem{lemma}[mcounter]{Lemma}
\newtheorem{remark}[mcounter]{Remark}
\newtheorem{notation}[mcounter]{Notation}
\title{A non-commutative algorithm for multiplying~$\matrixsize{5}{5}$ matrices using~$99$ multiplications}
\author{Alexandre.Sedoglavic@univ-lille.fr}
\begin{document}
\maketitle
\begin{abstract}
  We present a non-commutative algorithm for multiplying~$\matrixsize{5}{5}$ matrices using~$99$ multiplications. 
  This algorithm is a minor modification of Makarov's algorithm which
  exhibit the previous best known bound with~$100$ multiplications.
\end{abstract}
\section{Introduction}
In his seminal work~\cite{strassen:1969}, V.\ Strassen introduced a
non-commutative algorithm for multiplication of
two~$\matrixsize{2}{2}$ matrices using only~$7$ coefficient multiplications. 
Since, several algorithms where proposed for the product of small
matrices during the last~$40$ years (e.g.~\mbox{\cite{laderman:1976a,
    schachtel:1978, makarov:1986b}}).
In~$1987$, O.M.\ Makarov shown in~\cite{makarov:1987} that the product
of two~$\matrixsize{5}{5}$ matrices can be
done using~$100$ coefficient multiplications.
\par
In this note, we present a non-commutative~$M_{5\times 5 \times 5}$
performing the matrix product problem~${AB=C}$ expressed with
the following generic~$\matrixsize{5}{5}$ matrices:
\begin{equation}
\label{eq:M_5x5x5}
  M_{5\times 5 \times 5} : \!\left(\!%
    \begin{array}{ccc}
      {a_{11}}&\cdots&{a_{15}}\\
      \vdots& &\vdots\\
      {a_{51}}&\cdots&{a_{55}}
    \end{array}\!\right)
  \!\left(\!%
    \begin{array}{ccc}
      {b_{11}}&\cdots&{b_{15}}\\
      \vdots& &\vdots\\
      {b_{51}}&\cdots&{b_{55}}
    \end{array}\!\right)
  =\!\left(\!%
    \begin{array}{ccc}
      {c_{11}}&\cdots&{c_{15}}\\
      \vdots& &\vdots\\
      {c_{51}}&\cdots&{c_{55}}
    \end{array}\!\right);\\
\end{equation}
this algorithms requires~$99$ coefficient
multiplications.
Furthermore, we explain briefly how this result is derived from Makarov's original
algorithm and we conclude by some implications of this new upper bound.
\section{An algorithm for multiplying~$\matrixsize{5}{5}$ matrices}
The algorithm presented below was \emph{accidentally} obtained while implementing Makarov's algorithm in a computer algebra package devoted to matrix multiplication algorithms seen as geometric objects represented by tensors.  
In his original paper~\cite{makarov:1987}, O.M.\ Makarov decomposed the~$\matrixsize{5}{5}$ matrix multiplication algorithm as~$7$ others matrix multiplication algorithms using a variant of Strassen's algorithm~\cite{strassen:1969} and obtain an algorithm requiring~$101$ multiplications. 
Then, Makarov uses the particular form of one of the used sub-algorithms to obtain his algorithm that we are not be able to improved.  
But by explicitly implementing the intermediary algorithm
requiring~$101$ multiplications, the simplification procedure of our
computer algebra package produces automatically the following
evaluation scheme that requires only~$99$ coefficient multiplications:
\begin{align}
  \label{eq:M99_1}
  {m_{1}}&= \left({a_{51}}+{a_{35}}+{a_{45}}-{a_{55}} \right) {b_{54}},\\
  {m_{2}}&= \left(-{a_{12}}+{a_{14}} \right) {b_{21}},\\
  {m_{3}}&=-{a_{32}}\,{b_{23}},\\
  {m_{4}}&=-{a_{34}}\,{b_{43}},\\
  {m_{5}}&=\,{a_{14}}\, \left({b_{21}}+{b_{41}} \right)\!,\\
  {m_{6}}&= \left(-{a_{32}}+{a_{42}} \right) \left(
           {b_{14}}-{b_{24}}+{b_{34}}-{b_{44}} \right),\\
  {m_{7}}&= \left({a_{12}}+{a_{32}}-{a_{14}} \right) \left(
           {b_{21}}+{b_{41}}+{b_{43}} \right),\\
  {m_{8}}&=\,{a_{53}}\,{b_{35}}\,\ell,\\
  {m_{9}}&=\,{a_{51}}\,{b_{15}}\,\ell,\\
  {m_{10}}&= \left(-{a_{43}}+{a_{34}}-{a_{44}} \right) \left(
            {b_{21}}-{b_{41}}+{b_{12}}-{b_{22}}-{b_{32}}+{b_{42}} \right),\\
  {m_{11}}&= \left({a_{21}}-
            {a_{12}}+{a_{22}}+{a_{23}}-{a_{14}}+{a_{24}} \right) \left(
            -{b_{43}}-{b_{34}}
            +{b_{44}} \right),\\
  {m_{12}}&= \left({a_{32}}-{a_{42}}-{a_{14}}+{a_{24}} \right)
            \left({b_{12}}-{b_{22}}+{b_{14}}-{b_{24}}+{b_{34}}-{b_{44}} \right),\\
  {m_{13}} &= \left(-{a_{21}}+{a_{12}}-{a_{22}} \right) \left(
             -{b_{23}}+{b_{43}}-{b_{14}}
             +{b_{24}}+{b_{34}}-{b_{44}} \right),\\
  {m_{14}}&={a_{45}}\, \left({b_{32}}-{b_{42}}+{b_{52}} \right),\\
  {m_{15}}&= \left(
            {a_{12}} -{a_{21}}-{a_{22}}-{a_{43}}+{a_{34}}-{a_{44}} 
            \right) \left(
            \begin{array}{c}
              {b_{22}}-{b_{21}}-{b_{12}}  \\+{b_{44}} -{b_{43}}-{b_{34}}
            \end{array}
  \right),\\
  {m_{16}}&= \left(-{a_{41}}+{a_{32}}-{a_{42}}-{a_{43}}+{a_{34}
            }-{a_{44}} \right)  \left(-{b_{21}}-{b_{12}}+{b_{22}} \right),\\
  {m_{17}}&= \left(-{a_{23}}-{a_{25}} \right)  \left(-{b_{53}}+{b_{54}} \right),\\
  {m_{18}}&= \left({a_{43}}+{a_{44}} \right)  \left({b_{21}}-{b_{41}}-{b_{22}}+{b_{42}} \right),\\
  { m_{19}}&= \left(-{a_{21}}-{a_{22}}-{a_{23}}-{a_{24}} \right)
             \left(-{b_{43}}+{
             b_{44}} \right),\\
  {m_{20}}&= \left({a_{14}}-{a_{24}} \right) \left({b_{12}}-{
            b_{22}}+{b_{32}}-{b_{42}}+{b_{14}}-{b_{24}}+{b_{34}}-{b_{44}} \right),\\
  {m_{21} }&= \left({a_{21}}+{a_{22}} \right) \left(-{b_{23}}+{b_{43}}+{b_{24}}-{b_{44}}
             \right),\\
  {m_{22}}&= \left(-{a_{12}}-{a_{32}}+{a_{14}}+{a_{34}} \right)
            \left({
            b_{41}}+{b_{43}} \right),\\
  {m_{23}}&= \left({a_{12}}+{a_{32}} \right) \left({
            b_{21}}+{b_{41}}+{b_{23}}+{b_{43}} \right),\\
  {m_{24}}&= \left({a_{12}}-{a_{22}}-
            {a_{32}}+{a_{42}}-{a_{25}} \right)  \left({b_{12}}-{b_{22}} \right),\\
  {m_{25}}&= \left(-{a_{31}}+{a_{41}}-{a_{32}}+{a_{42}} \right)
            \left({b_{14}}+{b_{34}}
            \right),\\
  {m_{26}}&= \left(-{a_{41}}-{a_{43}} \right) \left(-{b_{13}}+{b_{23}}+
            {b_{14}}-{b_{24}} \right),\\
  {m_{27}}&={a_{35}}\, \left({b_{32}}+{b_{52}} \right),\\
  {m_{28}}&= \left(-{a_{54}}-{a_{45}} \right) \left({b_{32}}-{b_{42}}+{b_{52}}+
            {b_{54}} \right),\\
  {m_{29}}&= \left({a_{21}}+{a_{22}}+{a_{43}}+{a_{44}} \right)
            \left(-{b_{21}}+{b_{22}}-{b_{43}}+{b_{44}} \right),\\
  {m_{30}}&= \left({a_{41}}+
            {a_{42}}+{a_{43}}+{a_{44}} \right)  \left(-{b_{21}}+{b_{22}} \right),\\
  {m_{31}}&=
            -{a_{52}}\, \left({b_{25}}+{b_{45}} \right),\\
  {m_{32}}&= \left(\ell{a_{31}}+{a_{51} }-{\ell}^{2}{a_{35}} \right)
            \left({b_{11}}+{\frac{{b_{51}}}{\ell}}-\ell{b_{15}}
            \right),\\
  {m_{33}}&= \left({a_{23}}+{a_{25}}+{a_{45}} \right) \left({b_{53}}-{
            b_{54}}-{b_{35}}+{b_{45}} \right),\\
  {m_{34}}&= \left(\ell{a_{13}}+{a_{53}}-{\ell}^{2}{
            a_{15}} \right)  \left({b_{31}}+{\frac{{b_{51}}}{\ell}}-\ell{b_{35}} \right),\\
  {m_{35}}&= \left({a_{52}}-{a_{54}} \right) {b_{45}},\\
  {m_{36}}&= \left({a_{14}}-{a_{23}}+{a_{43}}-{a_{24}}-{a_{34}}+{a_{44}}
            \right) \left(\begin{array}{c}
              {b_{42} -{b_{41}}-{b_{32}}}\\ - {b_{43}}-{b_{34}}+{b_{44}}              
            \end{array}
  \right),\\
  {m_{37}}&=-{a_{52}}\, \left({b_{14}}-{
            b_{24}}+{b_{54}} \right),\\
  {m_{38}}&= \left({a_{21}}+{a_{41}}{a_{43}} \right) \left({b_{31}}-{b_{41}}-{b_{32}}+{b_{42}}-{b_{13}}+{b_{23}}+{b_{14}}-{b_{24}} \right),\\
  {m_{39}}&= \left({a_{53}}+{a_{54}}-{a_{35}} \right) \left({b_{32}
            }+{b_{52}}+{b_{54}} \right),\\
  {m_{40}}&= \left(-{\frac{{a_{11}}}{\ell}}-{\frac{{
            a_{51}}}{{\ell}^{2}}}+{a_{15}} \right) \left({b_{13}}+{\frac{{b_{53}}}{\ell}}-\ell{b_{15}}
            \right),\\
  {m_{41}}&= \left({a_{21}}-{a_{41}}-{a_{12}}+{a_{22}}+{a_{32}}-{a_{42} }
            \right) \left(\begin{array}{c}
              {b_{24}}-{b_{21}}-{b_{12}}\\ +{b_{22}}-{b_{23}}-{b_{14}}             
            \end{array}
  \right),\\
  {m_{42}}&={a_{25}}\, \left({b_{12}}-{b_{22}}+{b_{52}} \right),\\
  {m_{43} }&=-{a_{54}}\, \left({b_{32}}-{b_{42}}+{b_{52}}+{b_{34}}-{b_{44}}+{b_{54}}
             \right),\\
  {m_{44}}&= \left({a_{31}}-{a_{41}}+{a_{32}}-{a_{42}}-{a_{13}}+{a_{23}
            }-{a_{14}}+{a_{24}} \right)  \left({b_{12}}+{b_{14}}+{b_{34}} \right),\\
  {m_{45}}&= \left(-{a_{52}}-{a_{25}} \right)  \left({b_{12}}-{b_{22}}-{b_{54}} \right),\\
  { m_{46}}&= \left(\ell{a_{33}}+{a_{53}}-{\ell}^{2}{a_{35}} \right)
             \left({b_{33}}+{
             \frac{{b_{53}}}{\ell}}-\ell{b_{35}} \right),\\
  {m_{47}}&= \left({a_{52}}+{a_{14}}
            \right)  \left({b_{21}}-{b_{45}} \right),\\
  {m_{48}}&= \left({\frac{{a_{31}}}{\ell}} +{\frac{{a_{51}}}{{\ell}^{2}}} \right) \left({b_{11}}
            +{\frac{{b_{51}}}{\ell}}
            \right),\\
  {m_{49}}&={a_{45}}\, \left(-{b_{53}}+{b_{54}}-{b_{15}}+{b_{25}}+{
            b_{55}} \right),\\
  {m_{50}}&= \left(-\ell{a_{31}}+{\ell}^{2}{a_{35}} \right) \left({b_{11}}
            -\ell{b_{15}} \right),\\
  {m_{51}}&= \left({a_{21}}-{a_{45}} \right) \left(-{b_{51}}+
            {b_{52}}+{b_{15}}-{b_{25}} \right),\\
  {m_{52}}&=-{a_{21}}\, \left({b_{11}}-{
            b_{21}}-{b_{51}}-{b_{12}}+{b_{22}}+{b_{52}}-{b_{13}}+{b_{23}}+{b_{14}}-{b_{24}}
            \right),\\
  {m_{53}}&= \left({a_{21}}+{a_{25}} \right) \left({b_{51}}-{b_{52}}
            \right),\\
  {m_{54}}&= \left({\frac{{a_{13}}}{\ell}}+{\frac{{a_{53}}}{{\ell}^{2}}}
            \right)  \left({b_{31}}+{\frac{{b_{51}}}{\ell}} \right),\\
  {m_{55}}&= \left({a_{52}}
            -{a_{34}}-{a_{54}} \right)  \left({b_{23}}+{b_{25}}+{b_{45}} \right),\\
  {m_{56}}&=
            \left(-\ell{a_{13}}+{\ell}^{2}{a_{15}} \right)  \left({b_{31}}-\ell{b_{35}} \right),\\
  { m_{57}}&= \left({a_{23}}+{a_{43}}+{a_{25}}+{a_{45}} \right)
             \left({b_{35}}-{
             b_{45}} \right),\\
  {m_{58}}&= \left({a_{23}}-{a_{43}}+{a_{24}}-{a_{44}} \right)
            \left(-{b_{41}}+{b_{42}}-{b_{43}}+{b_{44}} \right),\\
  {m_{59}}&= \left(-{a_{41}}
            -{a_{45}} \right)  \left(-{b_{15}}+{b_{25}} \right),\\
  {m_{60}}&= \left({a_{13}}-{ a_{23}}+{a_{14}}-{a_{24}} \right)
            \left({b_{12}}+{b_{32}}+{b_{14}}+{b_{34}}
            \right),\\
  {m_{61}}&= \left({a_{53}}+{a_{54}} \right) \left({b_{32}}+{b_{52}}+{
            b_{34}}+{b_{54}} \right),\\
  {m_{62}}&= \left({a_{21}}+{a_{41}}+{a_{23}}+{a_{43}}
            \right)  \left({b_{31}}-{b_{41}}-{b_{32}}+{b_{42}} \right),\\
  {m_{63}}&= \left(-{ a_{21}}+{a_{41}}-{a_{22}}+{a_{42}} \right)
            \left(-{b_{21}}+{b_{22}}-{b_{23}}+
            {b_{24}} \right),\\
  {m_{64}}&= \left({\frac{{a_{11}}}{\ell}}+{\frac{{a_{51}}}{{\ell}^{2}}} \right)  \left({b_{13}}+{\frac{{b_{53}}}{\ell}} \right),\\
  {m_{65}}&= \left(-{ a_{32}}+{a_{42}}+{a_{34}}-{a_{44}}+{a_{54}}
            \right) \left({b_{34}}-{b_{44}}
            \right),\\
  {m_{66}}&= \left(-\ell{a_{11}}+{\ell}^{2}{a_{15}} \right) \left({b_{13}}-\ell{
            b_{15}} \right),\\
  {m_{67}}&={a_{15}}\, \left({b_{12}}+{b_{52}} \right),\\
  {m_{68}}&=
            \left({a_{51}}+{a_{52}} \right)  \left({b_{14}}+{b_{54}} \right),\\
  {m_{69}}&=
            \left(-\ell{a_{33}}+{\ell}^{2}{a_{35}} \right)  \left({b_{33}}-\ell{b_{35}} \right),\\
  { m_{70}}&= \left(-{a_{53}}-{a_{15}}-{a_{25}}+{a_{55}} \right)
             \left({b_{52}}+{
             b_{54}} \right),\\
  {m_{71}}&= \left({\frac{{a_{33}}}{\ell}}+{\frac{{a_{53}}}{{\ell}^{2}}
            } \right)  \left({b_{33}}+{\frac{{b_{53}}}{\ell}} \right),\\
  {m_{72}}&= \left(-{ a_{14}}+{a_{24}}+{a_{34}}-{a_{44}}-{a_{45}}
            \right) \left({b_{32}}-{b_{42}}
            \right),\\
  {m_{73}}&= \left({a_{11}}-{a_{21}}-{a_{31}}+{a_{41}}+{a_{12}}-{a_{22}
            }-{a_{32}}+{a_{42}}-{a_{15}} \right) {b_{12}},\\
  {m_{74}}&= \left({a_{12}}-{a_{22} }+{a_{52}}-{a_{14}}+{a_{24}}
            \right) \left({b_{12}}-{b_{22}}+{b_{14}}-{b_{24}
            } \right),\\
  {m_{75}}&= \left({a_{51}}+{a_{52}}-{a_{15}} \right) \left({b_{12}}-{
            b_{54}} \right),\\
  {m_{76}}&= \left({a_{21}}+{a_{41}} \right) \left(\begin{array}{c}
              {b_{11}}-{b_{21}}-{b_{31}}+{b_{41}}-{b_{12}} \\ +{b_{22}}+{b_{32}}-{b_{42}}-{b_{15}}+{b_{25}}               
            \end{array}
  \right),\\
  {m_{77}}&={a_{33}}\,{b_{31}},\\
  {m_{78}}&={a_{11}}\,{b_{11}},\\
  {m_{79}}&= \left(-{a_{13}}+{a_{23}}+{a_{33}}-{a_{43}}-{a_{14}}+{a_{24}}+{a_{34}}-{
            a_{44}}-{a_{35}} \right) {b_{32}},\\
  {m_{80}}&= \left({a_{14}}+{a_{54}} \right)
            \left({b_{41}}+{b_{45}} \right),\\
  {m_{81}}&= \left(-{a_{31}}-{\frac{{a_{51}}}{\ell}}
            -{a_{13}}-{\frac{{a_{53}}}{\ell}}+\ell{a_{15}}+\ell{a_{35}}+{a_{55}} \right) {b_{51}},\\
  {m_{82}}&={a_{23}}\, \left(-{b_{31}}+{b_{41}}+{b_{32}}-{b_{42}}+{b_{33}}-{
            b_{43}}-{b_{53}}-{b_{34}}+{b_{44}}+{b_{54}} \right),\\
  {m_{83}}&={a_{13}}\,{b_{33}},\\
  { m_{84}}&= \left(\begin{array}{c}
               {a_{11}}-{a_{21}}-{a_{51}}+{a_{12}}-{a_{22}}\\ -{a_{52}}-{a_{13}}+{a_{23}}-{a_{14}}+{a_{24}}                
             \end{array}
  \right)  \left({b_{12}}+{b_{14}} \right),\\
  {m_{85}}&= \left({a_{34}}+{a_{54}} \right) \left({b_{23}}+{b_{43}}+{b_{25}}+{b_{45}}
            \right),\\
  {m_{86}}&= \left(-{a_{11}}-{\frac{{a_{51}}}{\ell}}-{a_{33}}-{\frac{{a_{53}}}{\ell}}
            +\ell{a_{15}}+\ell{a_{35}}+{a_{55}} \right) {b_{53}},\\
  {m_{87}}&={a_{43}}\, \left(-{b_{13}}+{b_{23}}+{b_{33}}-{b_{43}}+{b_{14}}-{b_{24}}-{b_{34}}+{
            b_{44}}-{b_{35}}+{b_{45}} \right),\\
  {m_{88}}&={a_{31}}\,{b_{13}},\\
  {m_{89}}&={a_{15}}\, \left(-{b_{31}}-{\frac{{b_{51}}}{\ell}}-{b_{13}}-{\frac{{b_{53}}}{\ell}}+\ell{b_{15}}+
            \ell{b_{35}}+{b_{55}} \right),\\
  {m_{90}}&= \left(-{a_{31}}+{a_{41}}-{a_{32}}+{
            a_{42}}+{a_{33}}-{a_{43}}-{a_{53}}+{a_{34}}-{a_{44}}-{a_{54}} \right) {b_{34}},\\
  {m_{91}}&={a_{55}}\,{b_{55}},\\
  {m_{92}}&= \left({a_{43}}+{a_{44}} \right) {b_{45}},\\
  {m_{93}}&= \left({a_{41}}+{a_{42}} \right) {b_{25}},\\
  {m_{94}}&= \left({a_{32}}+{
            a_{52}}-{a_{34}}-{a_{54}} \right)  \left({b_{23}}+{b_{25}} \right),\\
  {m_{95}}&=-{ a_{35}}\, \left({b_{11}}+{\frac{{b_{51}}}{\ell}}+{b_{33}}+{\frac{{b_{53}}}{\ell}}-\ell{
            b_{15}}-\ell{b_{35}}-{b_{55}} \right),\\
  {m_{96}}&= \left({a_{23}}+{a_{24}} \right) {
            b_{45}},\\
  {m_{97}}&= \left({a_{21}}+{a_{22}} \right) {b_{25}},\\
  {m_{98}}&= \left({ a_{25}}+{a_{45}} \right) \left(-{b_{51}}+{b_{52}}-{b_{35}}+{b_{45}}+{b_{55}}
            \right),\\
  {m_{99}}&= \left({a_{12}}+{a_{52}} \right) \left({b_{21}}+{b_{25}} \right).
\end{align}
\begin{align}
  {c_{11}}&=-{\frac{{m_{8}}}{\ell}}-{m_{2}}+{m_{5}}+{m_{78}}+{m_{54}}\,\ell-{\frac{{
            m_{56}}}{\ell}}-{\frac{{m_{34}}}{\ell}},\\
  \begin{split}
    {c_{21}}&= {m_{10}}+{m_{11}}+{m_{12}}-{m_{2}}+{m_{5}}+{m_{6}}-{m_{52}}+{m_{53}}+{m_{62}} \\
    &+{m_{36}}-{m_{38}}+{m_{42}}+{m_{15}}+{m_{20}}+{m_{24}}-{m_{26}},
  \end{split}
  \\
  {c_{31}}&={m_{7}}-{\frac{{m_{9}}}{\ell}}+{m_{2}}+{
            m_{4}}+{m_{77}}-{\frac{{m_{50}}}{\ell}}-{\frac{{m_{32}}}{\ell}}+{m_{48}}\,\ell+{m_{22}}
            ,\\
  \begin{split}
    {c_{41}}&={m_{7}}+{m_{10}}+{m_{12}}+{m_{14}}+{m_{2}}+{m_{4}}+{m_{6}}+{m_{72}}
    +{m_{76}} \\
    &+{m_{51}}+{m_{52}}+{m_{59}}+{m_{38}}+{m_{16}}+{m_{20}}+{m_{22}}+{m_{26}},
  \end{split}\\
  \begin{split}
    {c_{51}}&=
    {m_{8}}+{m_{9}}-{m_{5}}+{m_{80}}+{m_{81}}+{m_{50}}\\
    &+{m_{56}}+{m_{32}}+{m_{34}}+{m_{35}}+{m_{47}},
  \end{split}\\
  \begin{split}
    {c_{12}}&={m_{10}}+{m_{11}}-{m_{2}}+{m_{5}}
    +{m_{67}}+{m_{73}}+{m_{58}}+{m_{60}}+{m_{36}} \\
    &+{m_{44}}+{m_{15}}+{m_{18}}+{ m_{19}}+{m_{25}}+{m_{29}},
  \end{split}\\
  \begin{split}
    {c_{22}}&={m_{10}}+{m_{11}}+{m_{12}}-{m_{2}}+{m_{5}}+{m_{6}}+{m_{58}}+{m_{36}}+{m_{42}} \\
    &+{m_{15}}+{m_{18}}+{m_{19}}+{m_{20}}+{m_{24} }+{m_{29}},
  \end{split}\\
  \begin{split}
    {c_{32}}&={m_{7}}+{m_{10}}+{m_{2}}+{m_{4}}+{m_{79}}+{m_{60}}+{
      m_{44}}+{m_{16}}+{m_{18}}\\
    &+{m_{22}}+{m_{25}}+{m_{27}}+{m_{30}},
  \end{split}\\
  \begin{split}
    {c_{42}}&={m_{7}}+{
      m_{10}}+{m_{12}}+{m_{14}}+{m_{2}}+{m_{4}}+{m_{6}}+{m_{72}}+{m_{16}}\\
    &+{m_{18}}+{m_{20}}+{m_{22}}+{m_{30}},
  \end{split}\\
  {c_{52}}&={m_{14}}+{m_{1}}+{m_{67}}+{m_{70}}+{
            m_{75}}+{m_{39}}+{m_{42}}+{m_{45}}+{m_{27}}+{m_{28}},\\
  {c_{13}}&=-{m_{7}}-{\frac{{
            m_{9}}}{\ell}}+{m_{3}}-{m_{5}}+{m_{83}}-{\frac
            {{m_{66}}}{\ell}}+{m_{64}}\,\ell+{m_{40}}
            \,\ell+{m_{23}},\\
  \begin{split}
    {c_{23}}&=-{m_{7}}-{m_{11}}-{m_{12}}+{m_{13}}+{m_{3}}-{m_{5}}-{
      m_{6}}+{m_{74}}+{m_{82}}\\
    &+{m_{62}}+{m_{37}}-{m_{38}}+{m_{45}}+{m_{17}}+{m_{23}}-{m_{24}}-{m_{26}},
  \end{split}\\
  {c_{33}}&=-{\frac{{m_{8}}}{\ell}}-{m_{3}}-{m_{4}}+{m_{88}}-{
            \frac{{m_{69}}}{\ell}}+{m_{71}}\,\ell-{\frac{{m_{46}}}{\ell}},\\
  \begin{split}
    {c_{43}}&={m_{13}}-{m_{14}}-{m_{3}}-{m_{4}}-{m_{6}}+{m_{87}}+{m_{57}}+{m_{65}}+{m_{33}}\\
    &+{m_{41}}+{m_{43}}+{m_{15}}-{m_{16}}-{m_{17}}+{m_{26}}-{m_{28}},
  \end{split}\\
  \begin{split}
    {c_{53}}&={m_{8}}+{m_{9}}+{m_{4}}+{m_{85}}+{m_{86}}+{m_{66}}\\
    &+{m_{69}}+{m_{55}}-{m_{40}}\,{\ell}^{2}+{m_{46}}+{m_{31}},
  \end{split}\\
  \begin{split}
    {c_{14}}&=-{m_{7}}-{m_{11}}+{m_{13}}+{m_{3}}-{m_{5}}+{m_{84}}+{
      m_{68}}-{m_{73}}\\
    &+{m_{75}}-{m_{44}}-{m_{19}}+{m_{21}}+{m_{23}}-{m_{25}},
  \end{split}\\
  \begin{split}
    {c_{24}}&=-
    {m_{7}}-{m_{11}}-{m_{12}}+{m_{13}}+{m_{3}}-{m_{5}}-{m_{6}}+{m_{74}}\\
    &+{m_{37}}+{m_{45}}-{m_{19}}+{m_{21}}+{m_{23}}-{m_{24}},
  \end{split}\\
  \begin{split}
    {c_{34}}&={m_{13}}-{m_{3}}-{m_{4}}+{m_{90}}+{m_{61}}+{m_{63}}-{m_{39}}+{m_{41}}\\
    &+{m_{15}}-{m_{16}}+{m_{21}}-{m_{25}}-{m_{27}}+{m_{29}}-{m_{30}},
  \end{split}\\
  \begin{split}
    {c_{44}}&={m_{13}}-{m_{14}}-{m_{3}}-{m_{4}}
    -{m_{6}}+{m_{63}}+{m_{65}}+{m_{41}}+{m_{43}}\\
    &+{m_{15}}-{m_{16}}+{m_{21}}-{m_{28}}+{m_{29}}-{m_{30}},
  \end{split}\\
  {c_{54}}&=-{m_{14}}-{m_{1}}+{m_{68}}+{m_{61}}+{m_{37}}-{
            m_{39}}+{m_{43}}-{m_{27}}-{m_{28}},\\
  \begin{split}
    {c_{15}}&=-{\frac
      {{m_{8}}}{{\ell}^{2}}}-{\frac{{m_{9}}}{{\ell}^{2}}}
    -{\frac{{m_{34}}}{{\ell}^{2}}}+{m_{2}}+{m_{89}}\\
    &+{m_{99}}+{m_{54}}+{m_{64}}+{m_{40}}-{m_{47}}+{m_{31}},
  \end{split}\\
  {c_{25}}&={m_{96}}+{m_{97}}+{m_{98}}-{m_{49}}+{m_{51}}+{m_{53}}-{m_{33}}+{m_{17}},\\
  \begin{split}
    {c_{35}}&=-{\frac{{m_{8}}}{{\ell}^{2}}}-{\frac{{m_{9}}}{{\ell}^{2}}}-{\frac{{m_{46}}}{{\ell}^{2}}}-{\frac{{m_{32}}}{{\ell}^{2}}}+{m_{3}}+{m_{94}}+{m_{95}}\\
    &+{m_{71}}-{m_{55}} +{m_{35}}+{m_{48}},
  \end{split}\\
  {c_{45}}&={m_{92}}+{m_{93}}+{m_{49}}+{m_{57}}+{m_{59}}+{m_{33}}-{m_{17}},\\
  \label{eq:M99_last}
  {c_{55}}&={\frac{{m_{8}}}{\ell}}+{\frac
            {{m_{9}}}{\ell}}+{m_{91}}-{m_{35}}-{m_{31}}.
\end{align}
\begin{remark}--- The free parameter~$\ell$ used in~(\ref{eq:M99_1})\,--\,(\ref{eq:M99_last})
  came from the utilisation of Winograd variant of Strassen algorithms (see~\cite{chatelin:1986a}) as presented in~\cite{sedoglavic:2017aa}.
\end{remark}
In the following section, we present explicitly the difference between the original Makarov's algorithm and the improved version presented here.
\section{Where does the improvement come from?}
Makarov's result is based on a divide-and-conquer strategy:
\begin{itemize}
\item The original problem is ``divided'' into~$7$ matrix multiplication
  subproblems using the Strassen's matrix multiplication
  algorithm~\cite{strassen:1969} (see Drevet et
  all~\cite{drevet:2011a} or Sedoglavic~\cite{Sedoglavic:2017ac} for a
detailled description of similar---but inequivalent---decompositions).
\item Each of these~$7$ subproblems could be handled by the more
  efficient known matrix multiplication algorithm adapted to its
  matrix sizes.  
These resolutions allow to ``conquer'' an algorithm solving the
original problem more efficiently then the trivial approach.
\end{itemize}
  \begin{notation}---
 In the sequel, we denote by~$\FMMA{a}{b}{c}{d}$ a matrix
 multiplication algorithm computing the product of a matrix of
 size~$\matrixsize{a}{b}$ by a matrix of size~$\matrixsize{b}{c}$
 using~$d$ coefficient multiplications. 
  \end{notation}
Makarov's algorithm~$\mathcal{M}$ relies on the following
subproblems:
  \begin{itemize}
  \item one~\href{http://cristal.univ-lille.fr/~sedoglav/FMM/2x2x2.html}{$\FMMA{2}{2}{2}{7}$} product~$M_{9}$ done by Strassen algorithm~$\mathcal{S}$~\cite{strassen:1969};
  \item one~\href{http://cristal.univ-lille.fr/~sedoglav/FMM/3x3x3.html}{$\FMMA{3}{3}{3}{23}$} product~$M_{4}$ done by Laderman algorithm~$\mathcal{L}$~\cite{laderman:1976a};
  \item one~{$\FMMA{2}{2}{3}{11}$} product~$M_{6}$ done by
    Hopcroft-Kerr algorithm~$\mathcal{K}$~\cite[Th~3]{hopcroft:1971}
    (a.k.a.\ basic use of Strassen algorithm);
  \item and four~\href{http://cristal.univ-lille.fr/~sedoglav/FMM/2x3x3.html}{$\FMMA{3}{3}{2}{15}$} products~${M_{3},M_{5},M_{1}}$
    and~$M_{2}$ done by
    Hopcroft-Kerr algorithms~$\mathcal{H}$~\cite{hopcroft:1971}
    (a.k.a.\ clearly not basic use of Strassen algorithm).
  \end{itemize}
Above, the indices~$i$ in~$M_{i}$ refer directly to
  Makarov's numeration in~\cite{makarov:1987} and 
the final algorithm~(\ref{eq:M_5x5x5}) could be obtained in a trilinear form by the sum:
\begin{equation}
  \label{eq:MakarovDecomposition}
\langle \mathcal{M} | M_{5\times 5 \times 5}\rangle=
\langle \mathcal{S} | M_{9}\rangle+
\langle \mathcal{L} | M_{4}\rangle+
\langle \mathcal{K} | M_{6}\rangle+
\sum_{i\in\{1,2,3,5\}}
\langle \mathcal{H} | M_{i}\rangle.
\end{equation}

The interested reader could found in~\cite{sedoglavic:2017aa} a brief description of
the framework (trilinear form, etc.) evoked above and in~\cite{Sedoglavic:FMMDB} the
complete description of the used algorithms. The---tedious---complete
presentation of these details is not necessary to expose the
improvement done to Makarov's algorithm.
\par
However, let us present with more  details Makarov's subproblem~\cite[$M_{1}$]{makarov:1987}
and~\cite[$M_{2}$]{makarov:1987} that are in our notations:
\begin{subequations}
  \begin{align}
    \label{eq:M1}
    \begin{split}
    M_{1} : & \left(%
    \begin{array}{ccc}
      a_{11}+a_{12}-a_{21}-a_{22} & a_{13}+a_{14}-a_{23}-a_{24}& a_{15} \\
      a_{31}+a_{32}-a_{41}-a_{42} & a_{33}+a_{34}-a_{43}-a_{44}& a_{35}\\
      a_{51}+a_{52} & a_{53}+a_{54}& a_{55}
    \end{array}
           \right) 
    \left(%
    \begin{array}{cc}
      b_{12} & b_{14}  \\
      b_{32} & b_{34} \\
      b_{52} & b_{54}
    \end{array}
         \right) \\ &=
    \left(%
    \begin{array}{cc}
      n_{12} & n_{14} \\
      n_{32} & n_{34}  \\
      \jmath c_{52} & \jmath c_{54}
    \end{array}
         \right) 
    \end{split}\\ 
    \begin{split}\label{eq:M2}
    M_{2}: & \left(%
    \begin{array}{ccc}
      a_{12}-a_{22} & a_{14}-a_{24}& a_{25} \\
      a_{32}-a_{42} & a_{34}-a_{44}& a_{45}\\
      -a_{52} & -a_{54}& 0
    \end{array}
           \right) 
    \left(%
    \begin{array}{cc}
      b_{12}-b_{22} & b_{14}-b_{24} \\
      b_{32}-b_{42}&  b_{34}-b_{44} \\
       b_{52}&  b_{54} 
    \end{array}
         \right) \\ &=
    \left(%
    \begin{array}{cc}
      n_{22} & n_{24} \\
      n_{42} & n_{44}  \\
     (1- \jmath) c_{52} & (1-\jmath) c_{54}
    \end{array}
         \right)
\end{split}
 \end{align}
\end{subequations}
The quantities~$n_{ij}$ stand for intermediate variables allowing the computation of the
wanted result~$C$ (similar to~$m_{i}$ in~(\ref{eq:M99_1})\,--\,(\ref{eq:M99_last})) 
and~$\jmath$ is a free parameters.
\par
As any other decomposition applied to this problem, the decomposition~(\ref{eq:MakarovDecomposition}) shows directly the following statement:
  \begin{lemma}---
Strassen, Laderman and Hopcroft-Kerr algorithms allow to
construct~$\FMMA{5}{5}{5}{101}$.    
  \end{lemma}
  \begin{remark}---
The second part of Makarov's paper use the fact that a
coefficient in~\cite[$M_{2}$]{makarov:1987} is~$0$ in order to show that this last
subproblem could be solved using~$14$ multiplications instead of~$15$
by avoiding a useless multiplication. 
Similarly, one can
obtain~$\FMMA{5}{5}{5}{100}$ by various decompositions (mainly based
on Winograd variant of Strassen algorithm) not necessarily equivalent
to Makarov's one.
  \end{remark}
However, Makarov's decomposition (in contrary to others decompositions
known by the author of this note)
is the only one where two subproblems share---without any further
manipulations like Pan's trilinear aggregations~\cite{pan:1984a}---some 
common terms in such a way that the total complexity is reduced.
\par
In fact, the part of the original problem corresponding to
subproblems~\cite[$M_{1}$]{makarov:1987} and~\cite[$M_{2}$]{makarov:1987}
could be computed using only~$28$ instead of expected~$30$ multiplications.
To show that very briefly, let us present---in trilinear form---the concerned terms of Hopcroft-Kerr
algorithm~$\mathcal{H}$:
\begin{subequations}
  \begin{align}
    \label{eq:HK_1}
    \begin{split}
    \left\langle\mathcal{H},M_{1}\right\rangle & = \cdots 
    \\ & +
{\jmath}\left({a_{55}}-{a_{51}}-{a_{52}}-{a_{35}}\right) {b_{54}} 
\left({c_{54}}-{c_{52}}\right) 
    \\ & + 
 {\jmath}\left({a_{55}}  -{a_{53}}-{a_{54}}-{a_{15}}\right)  
 \left({b_{52}}+{b_{54}} \right){c_{52}},
    \end{split}\\
    \label{eq:HK_2}
    \begin{split}
    \left\langle\mathcal{H},M_{2}\right\rangle & = \cdots     
    \\ &+
\left( 1-{\jmath} \right)
\left( {a_{52}}-{a_{45}} \right) {b_{54}}\left({c_{54}} -{c_{52}} \right)
    \\&+
    \left(1-{\jmath} \right)\left({a_{54}}-{a_{25}} \right)  \left({b_{52}}+{b_{54}} \right)
    {c_{52}}.
    \end{split}
  \end{align}
\end{subequations}
These four last trilinear terms could be factorize as two
trilinear terms:
\begin{equation}
    \label{eq:HK_1+HK_2}
    \begin{split}
    \left\langle\mathcal{H},M_{1}\right\rangle +
    \left\langle\mathcal{H},M_{2}\right\rangle = \cdots \\ +
\left(({a_{55}}-{a_{51}}-{a_{35}}) {\jmath}+
\left( 1-2{\jmath} \right) {a_{52}}-\left( 1-{\jmath} \right){a_{45}}
\right)
{b_{54}}\left({c_{54}}-{c_{52}}\right)
\\ + \left(({a_{55}}-{a_{53}}-{a_{15}}) {\jmath}+
\left( 1-2{\jmath} \right) {a_{54}}-\left( 1-{\jmath} \right){a_{25}}
\right)  \left( {b_{52}}+{b_{54}} \right) {c_{52}}.
    \end{split}
\end{equation}
This simplification was automatically produced by our pilote
computer algebra package and implies the new upper
bound~\href{http://cristal.univ-lille.fr/~sedoglav/FMM/5x5x5.html}{$\FMMA{5}{5}{5}{99}$}. 
We unfortunately do not have any geometric interpretation of this
simplification and thus, we do not know if it is possible to reproduce
it on other matrix multiplication algorithm obtained by a
divide-and-conquer process.
\section{Concluding remarks}
\begin{remark}--- The algorithm presented in this note could be used
  to improve slightly other matrix multiplication algorithm's bounds
  like~\href{http://cristal.univ-lille.fr/~sedoglav/FMM/10x10x10.html}{$\FMMA{10}{10}{10}{693}$} for example.
\end{remark}
\begin{remark}---
\label{rem:grouptheoretic}
 It is shown in~\cite{hart:2013a} that no group can realize~$\matrixsize{5}{5}$ matrix multiplication better then Makarov's algorithm using the group-theoretic approach of Cohn and Umans~\cite{cohn:2003a}.
  Hence, the algorithm presented in this note shows that this approach
  does not produce better algorithms then~$\FMMA{5}{5}{5}{99}$.
 The same assertion for~$\FMMA{3}{3}{3}{23}$ and~$\FMMA{4}{4}{4}{49}$ was proved in~\cite[Theorem~7.3]{Hedtke:2012aa}.
\end{remark}
\paragraph{Acknowledgment.}
The author would like to thank Alin Bostan for his always illuminating remarks and specially for stating Remark~\ref{rem:grouptheoretic}.
\bibliographystyle{acm}

\end{document}